\documentclass{iau} 
\usepackage{graphicx}

\title[The long-lived inner bar of NGC~1291] %% give here short title %%
{The long-lived inner bar of NGC1291}

\author[Jairo M\'endez-Abreu]   %% give here short author list %%
{Jairo M\'endez-Abreu$^{1,2}$, Adriana de Lorenzo-C\'aceres$^{1,2}$, and the TIMER team}
  \affiliation{$^{1}$Instituto de Astrof\'isica de Canarias, C/ V\'ia L\'actea s/n, E-38205 La Laguna, Tenerife, Spain\\
$^{2}$Departamento de Astrof\'isica, Universidad de La Laguna, E-38200 La Laguna, Tenerife, Spain\\ email: {\tt jairomendezabreu@gmail.com} \\[\affilskip]}

\pubyear{2019}
\volume{353}  %% insert here IAU Symposium No.
\jname{Galactic Dynamics in the Era of Large Surveys}
\editors{M. Valluri, \& J. A. Sellwood  }
\begin{document}

\maketitle

\begin{abstract}

  The question whether stellar bars are either transitory features or long-lived structures is still matter of debate. This problem is more acute for double-barred systems where even the formation of the inner bar remains a challenge for numerical studies. We present a thorough study of the central structures of the double-barred galaxy NGC~1291. We used a two-dimensional multi-component photometric decomposition performed on the 3.6μm images from S$^4$G, combined with both stellar kinematics and stellar population analysis carried out using integral field data from the MUSE TIMER project. We report on the discovery of the first Box-Peanut (B/P) structure in an inner bar detected in the face-on galaxy NGC~1291. The B/P structure is detected as bi-symmetric minima of the $h_4$ moment of the line-of-sight velocity distribution along the major axis of the inner bar, as expected  from  numerical  simulations. Our observations demonstrate that inner bars (similarly as outer bars) can suffer buckling instabilities, thus suggesting they can survive a long time after bar formation. The analysis of the star formation history for the structural components shaping the central regions of NGC~1291 also constrains the epoch of dynamical assembly of the inner bar, which took place $>$6.5 Gyr ago for NGC~1291. Our results imply that the inner bar of NGC~1291 is a long-lived structure.
 
\keywords{galaxies: individual: NGC~1291 -- galaxies: structure -- galaxies: kinematics and dynamics -- galaxies: evolution -- techniques: spectroscopic}
\end{abstract}

\firstsection % if your document starts with a section,
\section{Introduction}

\cite{devaucouleurs75}   described  a   rarity  in   the  center   of NGC~1291. He detected, for the first  time, an inner bar which follows the same lens-bar-nucleus pattern of the outer bar.  Nowadays, we know that  bars within  bars are  not an  oddity, with observations suggesting that $\sim30$\% of all  barred galaxies host an inner bar (\cite[Erwin 2004]{erwin04}). The importance of inner bars is not restricted to their high incidence. They are thought to be an efficient  mechanism for transporting gas to  the galaxy central regions, possibly fueling active galactic nuclei (AGN; Shlosman et al. 1990)  and affecting  the formation of new stellar structures (\cite[de Lorenzo-C\'aceres et al. 2012, 2013]{delorenzocaceres12,delorenzocaceres13}). Still, little is known about  the origin  of inner  bars. Now, 40  years later,  NGC~1291 strikes  again  providing  a  new  piece of  evidence  to  unveil  the formation of the central regions of galaxies.

\section{Detection of a Box/Peanut structure in the inner bar of NGC~1291}

%-----------------------------------------------------------------------
\begin{figure}[!t]
% \vspace*{-2.0 cm}
\begin{center}
 \includegraphics[bb= 120 -300 490 800,angle=90,width=18.5cm]{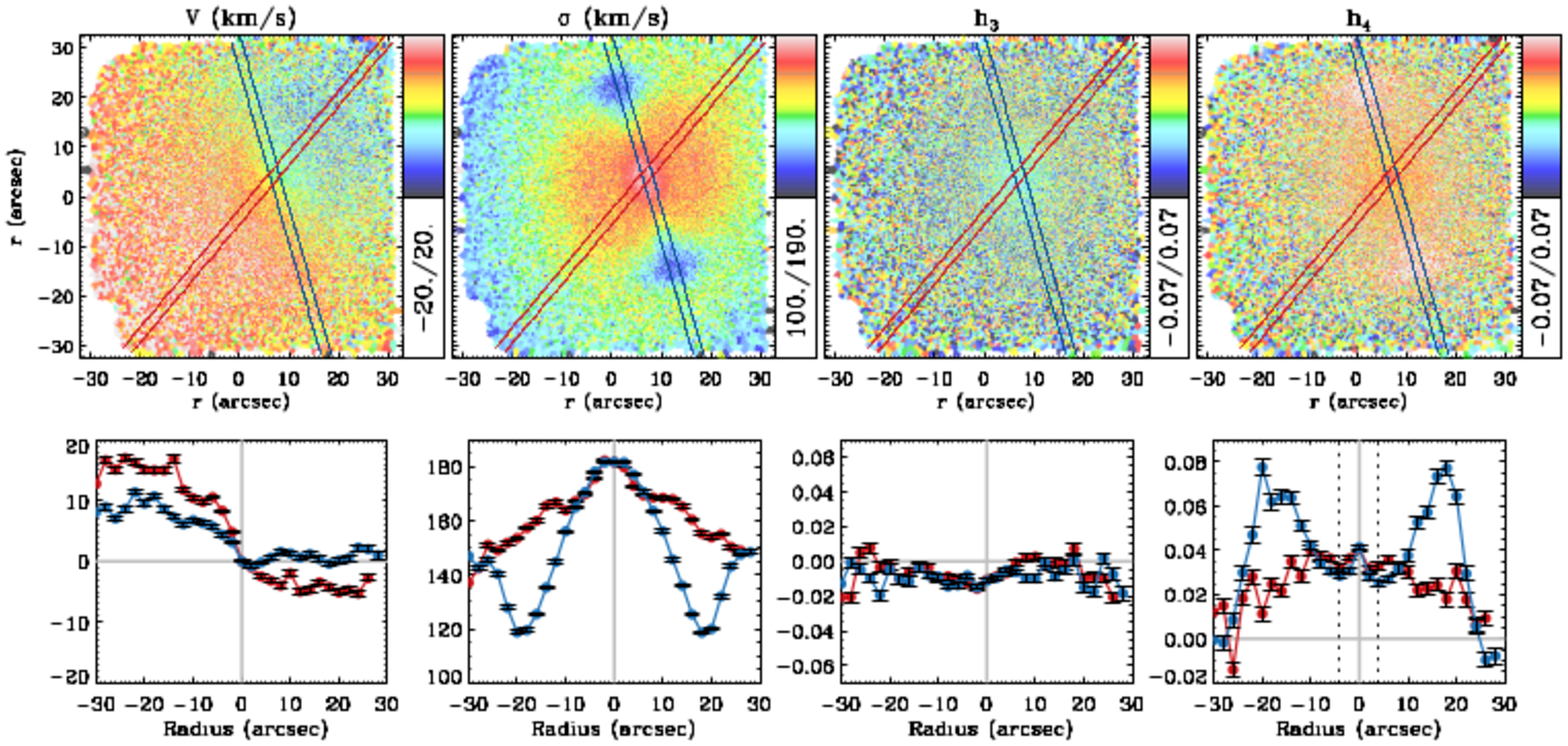} 
% \vspace*{-1.0 cm}
 \caption{Stellar  kinematic maps  (top rows)  and radial  profiles
      (bottom  rows)  obtained from  the  MUSE  TIMER observations  of
      NGC~1291.  From  left to  right: velocity,  velocity dispersion,
      $h_3$, and  $h_4$ moments.   The radial  profiles in  the bottom
      panels were extracted along 2'' width pseudo-slits (shown
      in the  maps) for both the  inner bar major axis  (blue) and the
      galaxy major axis  (red). The positions of the  $h_4$ minima are
      shown with dashed lines in the bottom right panel. North is
        up and East is left. \cite{mendezabreu19}.}
   \label{fig1}
\end{center}
\end{figure}
%-----------------------------------------------------------------------

The measured stellar kinematic maps for NGC~1291 are shown in Fig.~\ref{fig1}. The  rightmost panels show  the $h_4$  spatial distribution and radial profiles (along the inner bar and galaxy disc) of  NGC~1291. Clear, symmetric, double $h_4$ minima are seen along the major axis of the inner bar.  These minima are neither observed along the galaxy nor the main bar major axis indicating that they are related to the presence of the inner bar. The double minima occur at a projected distance of $\sim 4''$, which corresponds to 26\% of the inner bar effective radius. Following these  minima, the radial  profile of $h_4$ along the major axis of the bar grows until it reaches two maxima at radii comparable to the inner bar radius. These maxima are not part of a positive $h_4$ ring around the inner bar, but they nicely correspond to the $\sigma-$hollows observed in the velocity dispersion profile. The $\sigma-$hollows are a  kinematic confirmation for the  presence of an inner bar \cite[(see De Lorenzo-C\'aceres et al. 2008)]{delorenzocaceres08}, and are also not observed along the major axis of the galaxy.

As  shown by  numerical simulations  of single-barred  galaxies (\cite[Debattista et al. 2005, Ianuzzi \& Athanassoula 2015]{debattista05,iannuzzi15}),  bi-symmetric  minima  in  the  $h_4$ profile along the bar major axis indicate the presence of a vertically extended B/P structure in the bar. This kinematic criterion to detect B/P in face-on galaxies has been previously confirmed in main bars (\cite[Mendez-Abreu et al. 2008b, 2014]{mendezabreu08b,mendezabreu14}), {\em but this  is the first time a B/P  is detected in a  face-on inner bar}. Hints of multiple B/P   in   edge-on   galaxies have also been discussed in \cite{ciamburgraham16}. These results strongly suggest that inner and outer bars in double-barred galaxies are governed by the same physical processes.

\section{Stellar populations analysis of the central structures of NGC1291}

%-----------------------------------------------------------------------
\begin{figure}[!t]
% \vspace*{-2.0 cm}
\begin{center}
 \includegraphics[bb=10 250 600 600,width=13cm]{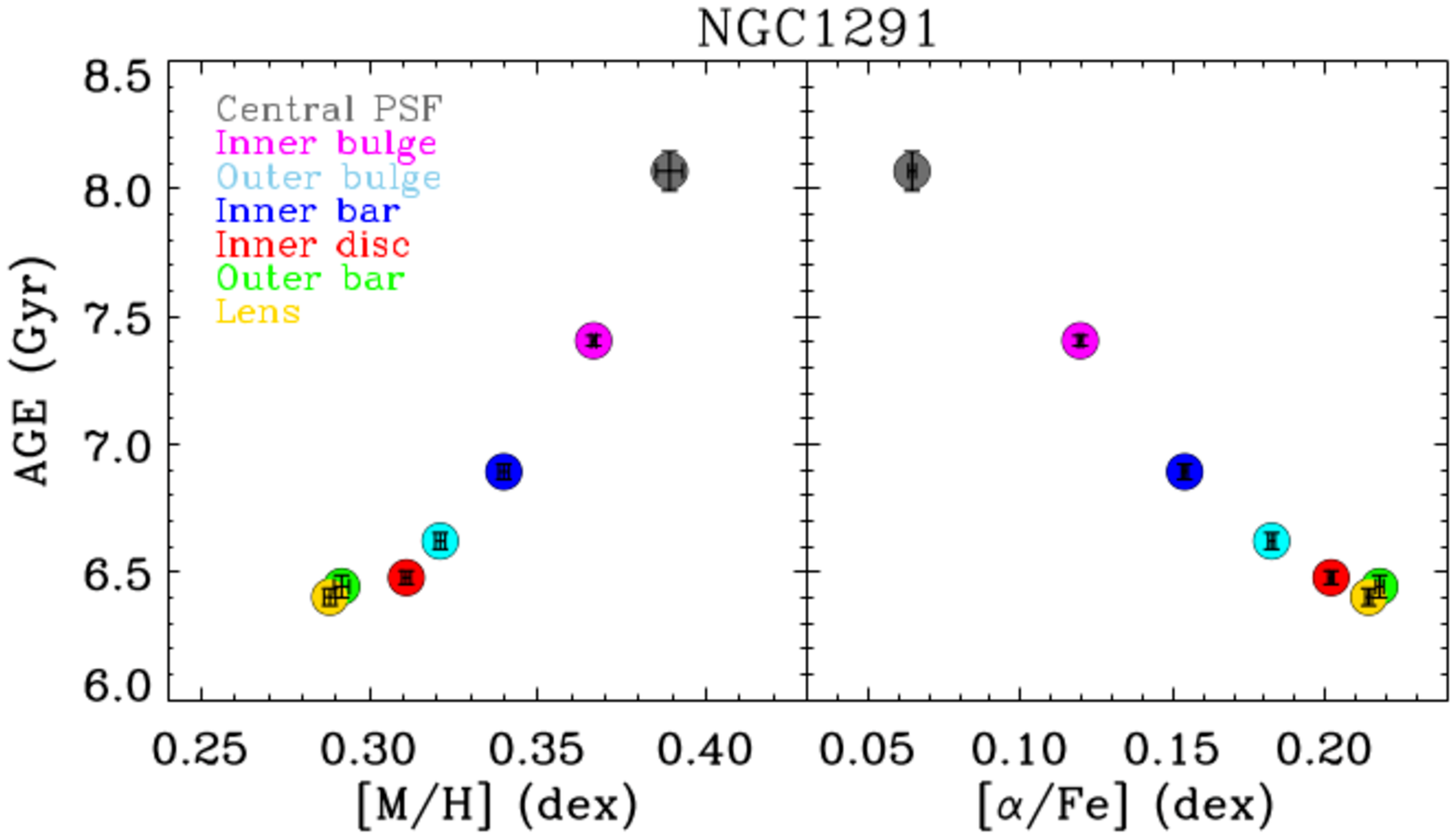} 
% \vspace*{-1.0 cm}
 \caption{Mean luminosity-weighted age versus metallicity (left panel) and [$\alpha$/Fe] (right panels) within the spatial regions where each structural component of NGC~1291 dominates the light: inner bulge (magenta), outer bulge (cyan), inner disc (red), inner bar (blue), outer bar (green), and lens (yellow). Due to possible resolution effects, the central PSF is considered separately (grey). \cite{delorenzocaceres19}.}
   \label{fig2}
\end{center}
\end{figure}
%-----------------------------------------------------------------------

The analysis of the stellar populations of the TIMER galaxies is extensively described in \cite[Gadotti et al. (2015, 2019)]{gadotti15,gadotti19}. Since we aim at discussing the structural assembly history of double-barred systems using the star formation history (SFH) results, our strategy consists in comparing the measurements within inner bars with those within other structural components. For this purpose, we average the relative contributions of the different stellar populations over apertures where the light contribution of each structure is mostly dominant. The relative light contribution from each structural component is computed using 2D photometric decompositions carried out with the GASP2D code (M\'endez-Abreu et al 2008a, 2017) on the 3.6$\mu$m images from S$^4$G, and projecting it onto a MUSE FoV.

Figure 2 shows the mean luminosity-weighted ages, metallicities, and [$\alpha$/Fe] linearly averaged within the regions of predominance of each individual structure. An inside-out gradient is clearly seen in Fig. 2, not only in age, but also in metallicity and [$\alpha$/Fe]. The centre (represented by the central PSF and the inner bulge) is the oldest, more metal-rich, and less $\alpha$-enhanced part of the galaxy. The remaining structures (outer bulge, inner bar, inner disc, outer bar, and lens) show old and particularly similar ages: values expand from 6.4 to 6.9 Gyr, and also the metallicity and [$\alpha$/Fe] cover short ranges. This result points towards a rapid consumption of most of the gas available for triggering significant star formation, so most of the stars were formed at an early stage.

\section{Discussion and Conclusions}

%-----------------------------------------------------------------------
\begin{figure}[!t]
% \vspace*{-2.0 cm}
\begin{center}
 \includegraphics[bb=10 200 600 630,width=11.6cm]{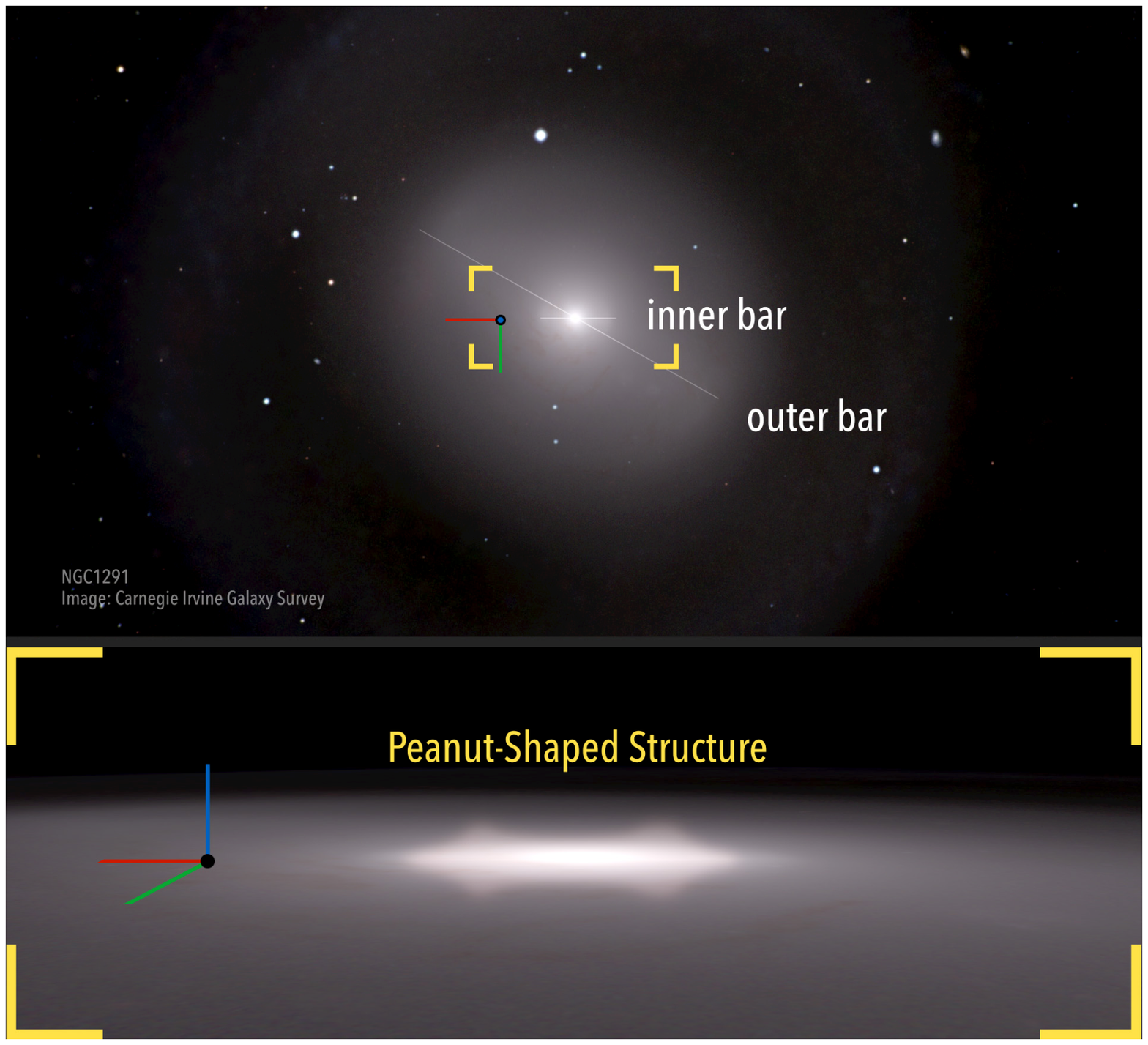} 
% \vspace*{-1.0 cm}
 \caption{Image of NGC1291 obtained from the Carnegie Irvine Galaxy Survey (Upper panel). The position of the outer and inner bars are highlihted with thin lines. The zoom-in region represented in the bottom panel is shown with green corners. Artistic representation of the B/P structure if NGC~1291 were seen edge-on (bottom panel). Credit: Gabriel P\'erez - IAC}
   \label{fig3}
\end{center}
\end{figure}
%-----------------------------------------------------------------------

Within an scenario where inner bars form dynamically out of inner discs (as large-scale bars), the bar structures in NGC~1291 were born from dynamical instabilities of the outer and inner disc. The inner disc was secularly formed after gas inflow along the outer bar. This process does involve star formation and the fact that the inner disc has a similar age as the outer galaxy regions indicate that the process of both outer and inner bar formation was fast. This explains that no rejuvenation of the inner disc with respect to the outer bar is noticeable, but the inner disc is slightly more metal-rich and $\alpha$-enhanced.

Once the inner disc is formed, the dynamical assembly of the inner bar happened at a later stage from the same stellar content. The fact that the inner bar is older than the inner disc is explained by invoking some star formation happening in the inner disc (as in the outer regions) after the inner bar was already assembled. This is not surprising: the shear exerted by the inner bar prevents star formation inside it, while the gas still present in the galaxy may continue forming stars outside.

Important conclusions are inferred from this assembly history, as it implies this inner bar is a long-lived structure. This conclusion is further reinforced by the presence of a box/peanut associated to the inner bar of NGC~1291 (see Fig. 3). The transient vs. long-living nature of inner bars has been a matter of debate for long time, but we show here compelling evidences for a long-lived inner bar in NGC~1291.

\end{document}